\documentclass[12pt]{article}
\usepackage{graphicx}
\usepackage{amssymb}
\usepackage{amsmath}
\usepackage{cite}
\newcommand{\norm}[1]{\left\Vert#1\right\Vert}

\newcommand{\A}{\mathcal{A}}
 \newcommand{\sgn}{\operatorname{sgn}}
\numberwithin{equation}{section}

\def\BR{{\mathbb R}}

\def\ve{{\varepsilon}}
\def\nn{{\nonumber}}

\newtheorem{Pa}{Paper}[section]
\newtheorem{Tm}[Pa]{{\bf Theorem}}

\newtheorem{Cy}[Pa]{{\bf Corollary}}
\newtheorem{Rk}[Pa]{{\bf Remark}}

\newtheorem{Ee}[Pa]{{\bf Example}}
\newtheorem{Dn}[Pa]{{\bf Definition}}
\newtheorem{Pn}[Pa]{{\bf Proposition}}
\usepackage{amsmath}

\title{Heisenberg's S-matrix program and Feynman's  divergence problem}

\author{Lev Sakhnovich}

\date{}

\begin{document}

\maketitle


\noindent \emph{Retired from Courant Institute, NY,}\\
\noindent \emph{99 Cove ave. Milford, CT 06461, USA, }\\
\noindent E-mail: lsakhnovich@gmail.com\\

 \noindent\textbf{MSC (2010):} Primary 81T15, Secondary 34L25, 81Q05,  81Q30.\\

\noindent {\bf Keywords:}  Generalized wave operator, deviation factor, secondary  generalized    scattering operator, infrared divergence, 
ultraviolet divergence.
\begin{abstract}
In the present article, we assume that   the first approximation of the scattering operator is given
and that it has the logarithmic divergence. This first approximation allows us to construct
the so called deviation factor. Using the deviation factor, we regularize all terms of the scattering operator's approximations.
The infrared and ultraviolet cases as well as concrete examples are considered.
Thus, for a wide range of cases, we provide a positive answer to the well-known problem of  J. R. Oppenheimer regarding scattering operators in QED:
``Can the procedure be freed of the expansion in $\varepsilon$ and carried out rigorously?"
\end{abstract}
\section{Introduction} \label{intro}
 In the present article, we assume that   the first approximation of the scattering operator is given
and it has the logarithmic divergence \cite{AB}.  The so called deviation factor 
was constructed using this first approximation in our papers \cite{Sakh10, Sak} (see Appendix \ref{ApA} for further details). Using the deviation factor, we regularize all the terms of the scattering operator's series representation.
\begin{Rk}\label{Remark 1.1} 
As was already stated by R. Feynman himself \cite{Fey1}  and remained true afterwards ``a strict basis for the rules of convergence is not known".
 A closely related basic question was formulated by J.R. Oppenheimer in the following form \cite{Opp}:\\
 {\rm  ``Can the procedure be freed of the expansion in $\varepsilon$ and carried out rigorously?"}
 In our article, we  give a   positive answer to this question.
 \end{Rk}
 In particular, a mathematical justification of the Feynman's theory \cite{AB, Dy} is presented for the important case of the  logarithmic divergence.

We start with  the self-adjoint operators $A,\,A_0,\,A_1$ which are defined in Hilbert space $H$ and are connected by the relation
\begin{equation} A=A_0+{\varepsilon}A_1, \label{1.1}\end{equation}
Here, $A$ is a perturbed operator, $A_0$ is an unperturbed operator, and $A_1$ is a perturbation operator.
In section \ref{sec2} on Coulomb type potentials, we consider  concrete  examples of $A,\,A_0,\,A_1$, where deviation factor theory is applied  to the radial  Schr\"odinger equation  and  to the radial Dirac equation  with Coulomb type potentials, as well as to the
Friedrichs model with a discontinuous kernel.
These examples provide heuristics for our further research.
In the main   scheme  and the corresponding proofs, these examples are not used.

Section \ref{sec3} is dedicated to the  {\it secondary generalized    scattering} and  perturbation operators.
(Note that {\it generalized    scattering} operators used, for instance, in \cite{Sakh3, Sakh1, Sakh10, Sakh10+, Sak} are introduced
in appendix \ref{ApA}.) At the beginning of section \ref{sec3} ,
 we consider 
 the operator function
\begin{equation} S(t,\tau,\varepsilon)=\exp(i t A_0)\exp(-i tA)\exp(i\tau A))\exp(-i\tau A_0), \label{1.2}\end{equation}
which is closely related to the scattering operator $S(A,A_0)$ (see \cite{Sakh1, Sakh2}). 
The operators  $A$ and $A_0$ above satisfy  \eqref{1.1}. Hence, it follows from \eqref{1.2} that
\begin{equation}\frac{\partial}{\partial{t}}S(t,\tau,\varepsilon)=-i{\varepsilon}V(t)S(t,\tau,\varepsilon),\quad
\frac{\partial}{\partial{\tau}}S(t,\tau,\varepsilon)=i{\varepsilon}S(t,\tau,\varepsilon)V(\tau), \label{1.3}\end{equation}
where
\begin{equation} V(t)=\exp(itA_0)A_1\exp(-itA_0),\quad S(t,t,\varepsilon)=I,\label{1.4}\end{equation}
and $I$ is the identity operator.
According to \eqref{1.4},  the  self-adjoint operator function $V(t)$  may be considered (at each $t$) as a  special representation of the perturbation operator $A_1$.
\begin{Dn}\label{Definition 1.2}The self-adjoint operator function $V(t)$  is  called the perturbation operator function.\end{Dn}
Further in section \ref{sec3}, we investigate the generalized systems \eqref{1.3}, that is, we do not suppose that $V(t)$ has the form \eqref{1.4}.
We define the operator function $V(t)$ using the first approximation of the scattering operator (see \eqref{3.24}).

We do not use operators  $A$, $A_1$ and $A_0$. These operators  are not given (and, may be, they do not exist at all).
\emph{Hence, the  presented in this paper results  follow the suggestions of  Heisenberg's $S$-program \cite{Hei}.} 
In this way, our approach to quantum 
electrodynamics here is similar to the approach to classical mechanics from our paper \cite{Sakh6}.

 In sections \ref{sec4} (on ultraviolet divergence) and \ref{sec5} (on concrete examples), we apply the results of section 
 \ref{sec3} to the well-known divergence problems in quantum electrodynamics.
 
The last section ``Conclusion" is devoted to the interpretation of the results obtained in this article.

In  appendix \ref{ApA} we give a definition of  the deviation factors when the operators $A,\, A_0, \, A_1$  are known.
 Then, we generalize this notion   for the case where the operators  $A,\, A_0 , \, A_1$  are either unknown or do not exist.
 
 In appendix \ref{ApB} the concept and an important property of the multiplicative integrals  are shortly introduced \cite{DF}.
 The case of the so called ``time infinity" in our theory is considered in Appendix \ref{ApC}.
 \begin{Rk}\label{Remark 1.3} The well-known divergence problems
 in quantum electrodynamics are old, classical  and very important. They have been studied
 by many outstanding physicists. These divergencies appear when the small parameter $\varepsilon$ expansions  are considered.
 We study the equation \eqref{1.3} without expanding its solution into a power series.\end{Rk}

\section{Coulomb type potentials: classical cases}\label{sec2}
{\bf 1.}  In this section, we study important cases of the explicit expressions for the deviation factors $W_0(t)$.
The deviation factors as well as the generalized wave and scattering operators are introduced in appendix \ref{ApA}
(see Definitions \ref{Definition 7.1} and \ref{Definition 7.2}).
\begin{Ee}\label{Example 2.1} Let us consider the radial  Schr\"odinger operator
\begin{equation}Af=-\frac{d^{2}}{dr^{2}}f+\left(\frac{\ell(\ell+1)}{r^2}-\frac{2{\varepsilon}z}{r}+{\varepsilon}q(r)\right)f,
\quad z=\overline{z} \label{2.1}\end{equation}
with Coulomb type potentials $2{\varepsilon}z/r-{\varepsilon}q(r)$, where ${\varepsilon}$ is a small parameter, $q(r)$ satisfies the condition $q(r)=\overline{q(r)}$ and
\begin{equation} \int_{0}^{1}|q^{2}|r^{2}dr<\infty,\quad  \int_{1}^{\infty}|q^{n}(r)|dr<\infty, \quad (n=1,\, 2).\label{2.2}\end{equation}
\end{Ee}
Here, $A$ acts in $H=L^{2}(0,+\infty)$. The operators ${A}_{0}$ and ${A}_{1}$ in this example have the form:
\begin{equation}{A}_{0}f=-\frac{d^{2}}{dr^{2}}f+\frac{\ell(\ell+1)}{r^2}f,\quad
 \quad {A}_{1}f=\left(-\frac{2z}{r}+q(r)\right)f.
\label{2.3}\end{equation}
The following statement was proved in \cite[section 6]{Sakh2}:
\begin{Tm}\label{Theorem 2.2}Suppose that $q(r)$ satisfies  \eqref{2.2}. Then, the generalized wave operators $W_{\pm}({A},{A}_{0})$  and
the generalized scattering operator $S({A},{A}_{0})$ exist. The corresponding deviation factor has the form
\begin{equation}W_{0}(t)=|t|^{i \sgn(t) \varepsilon z A_0^{-1/2}}.\label{2.4}\end{equation}
\end{Tm}
A special case of  Theorem \ref{Theorem 2.2}, where  $q(r) \equiv 0$, was proved  by J.G. Dollard \cite{Do}.

2. Let us consider the radial Dirac  equation.
\begin{Ee}\label{Example 2.3} The  radial  Dirac equation has the form:
\begin{align}&
\left( \frac{d}{dr}+\frac{k}{r}\right) f(r)-\big(\lambda+m+\frac{{\varepsilon}z}{r}-{\varepsilon}q(r)\big) g(r)=0,
\label{2.5}
\\ &
\left( \frac{d}{dr}-\frac{k}{r}\right) g(r)+\big(\lambda-m+\frac{{\varepsilon}z}{r}-{\varepsilon}q(r)\big) f(r)=0,
\label{2.6}\end{align}
where $\lambda = \overline{\lambda},\, m>0,\, k=\overline{k}$,  $\varepsilon$ is a small parameter and
\begin{equation}\quad z=\overline{z}\not=0,\quad |k|>{\varepsilon}|z|.\label{2.7}\end{equation}\end{Ee}
The corresponding Dirac operator $ {A}$  acts in the space $L_{2}^{2}(0,\infty)$ and has the form
\begin{equation}{A}F=\left(J_{1}\frac{d}{dr}+J_{2}\frac{k}{r}+mJ_{3}+{\varepsilon}v(r)\right)F,
\quad F(r)=\begin{bmatrix} f(r) \\ g(r)\end{bmatrix},
\label{2.8}\end{equation}
where $v(r)=\frac{z}{r}-q(r)$, $\,\,I_2$ is the $2\times 2$ identity matrix, and
\begin{equation}J_{1}=\left[\begin{array}{cc}
                        0 & 1 \\
                        -1 & 0
                      \end{array}\right],\quad J_{2}=\left[\begin{array}{cc}
                        0 & 1 \\
                        1 & 0
                      \end{array}\right],\quad
 J_{3}=\left[\begin{array}{cc}
                        -1 & 0 \\
                        0 & 1
\end{array}\right].\label{2.9}\end{equation}
Here, the unperturbed operator $A_0$ and the perturbation operator $A_1$ are given by the formulas: 
\begin{equation}{A}_{0}F=\left(J_{1}\frac{d}{dr}+J_{2}\frac{k}{r}++mJ_{3}\right)F,\quad A_{1}F=v(r)F.\label{2.10}\end{equation}
We assume that the following conditions are valid:\begin{equation}\int_{0}^{a}|rq(r)|dr+\int_{a}^{\infty}|q(r)|dr<\infty,\quad q(r)=\overline{q(r)}.\label{2.11}\end{equation}
The following statement is proved in our paper \cite{Sakh1}:
\begin{Tm}\label{Theorem 2.4}Suppose that $q(r)$ satisfies  \eqref{2.2}. Then, the generalized wave operators $W_{\pm}({A},{A}_{0})$  and
the generalized scattering operator $S({A},{A}_{0})$ exist. The corresponding deviation factor has the form
\begin{equation}W_{0}(t)=|t|^{i \sgn(t){\varepsilon}{z}\phi({A}_{0})},
\label{2.12}\end{equation}
where
\begin{equation}\phi({A}_{0})={A}_{0}({A}_{0}^2-m^2 I)^{-1/2}.\label{2.13}\end{equation}
\end{Tm}

{\bf 3.} Next, we consider Friedrichs model \cite{Fr}.
\begin{Ee}\label{Example 2.5} In this model, we have
\begin{equation}Af=xf(x)+\varepsilon\int_{a}^{b}f(s)K(x,s)ds,\quad A_{0}f=xf(x),\label{2.14}\end{equation}
where
 $f{\in}L_{n}^{2}[a,b]\,\, (-\infty<a<b<+\infty)$ and $ K(x,s)$ is  an  $ n{\times}n $ matrix valued function  (matrix function)  such that
$K(x,s)=K^{*}(s,x)$.
\end{Ee}
We assume below that the kernel $ K(x,s)$ is continuous at all points $x \not= s$ and has a discontinuity
of the first kind for  $x=s$. Since $K(x,s)=K^{*}(s,x)$, the matrix function
\begin{equation}P(x)=i[K(x,x+0)-K(x,x-0)]\label{2.15}\end{equation}
is self-adjoint. We suppose that
\begin{equation}P(x) \geq 0.\label{2.16}\end{equation}
Let us introduce the new integral kernels
\begin{equation}K_{0}(x,s)=(i/2)\sgn(x-s)P^{1/2}(s)P^{1/2}(x) \label{2.17}\end{equation}
 and
\begin{equation}K_{1}(x,s)=K(x,s)-K_{0}(x,s). \label{2.18}\end{equation}
In \cite[section 2]{Sakh2}, we  proved the following statement:
\begin{Tm}\label{Theorem 2.6}Suppose $P$ and $K_1$ satisfy H\"{o}lder conditions
\begin{equation}
\norm{P(x_2)-P(x_1)}{\leq}M|x_2-x_1|^{\alpha},\quad
\norm{K_1(x,s_2)-K_1(x,s_1)}{\leq}M|s_2-s_1|^{\alpha_1}\label{2.19}\end{equation}
 for some $M>0$, $\alpha>0$ and  $\alpha_1>1/2.$ Then, the generalized wave operators
$W_{\pm}(A,A_0)$ exist  and the deviation factor has the form
\begin{equation}W_{0}(t)f=|t|^{i \varepsilon P(x)}f\label{2.20}\end{equation}
where $f{\in}L_{n}^{2}[a,b] \,\, (-\infty<a<b<+\infty)$.\end{Tm}

4. Finally, consider the differential operator $A$ below (acting in the space $H= L^2(-\infty, +\infty)$).
\begin{Ee}\label{Example 2.7}\begin{equation}{A}f=-\frac{d^{2}}{dx^{2}}f+\left(
\frac{2{\varepsilon}zx}{x^2+a^2}+{\varepsilon}q(x)\right)f, \label{2.21}\end{equation}
where
\begin{equation} q(x)=\overline{q(x)},\quad q(x)\in L(-\infty,+\infty),\quad a>0.\label{2.22}\end{equation}
\end{Ee}
The corresponding operator $A_0$ has the form
\begin{equation}{A}_{0}f=-\frac{d^{2}}{dx^{2}}f.\label{2.23}\end{equation}
The following statement is proved in  \cite[section 4]{Sakh2}.
\begin{Tm}\label{Theorem 2.8}Suppose that $q(x)$ satisfies  \eqref{2.22}. Then, the generalized wave operators $W_{\pm}({A},{A}_{0})$  and
the generalized scattering operator $S({A},{A}_{0})$ exist. The corresponding deviation factor has the form
\begin{equation}W_{0}(t)f=\frac{1}{\sqrt{2\pi}}\int_{-\infty}^{+\infty}e^{-ikt}|t|^{z{\varepsilon}/k}g(k)dk, \label{2.25}\end{equation}
where $f$ is the Fourier transformation of $g:$
\begin{equation} f(x)=\frac{1}{\sqrt{2\pi}}\int_{-\infty}^{+\infty}e^{-ikt}g(k)dk. \label{2.24}\end{equation}
\end{Tm}
In the coordinate space, relation  \eqref{2.25} may be rewritten in the form
\begin{equation} W_{0}(t)f=|t|^{iz\varepsilon B_0^{-1}}f,\quad B_0f=i\frac{d}{dx}f. \label{2.26}\end{equation}
\section{Secondary  generalized    scattering  \\  and perturbation operators, \\ infrared divergence}\label{sec3}
{\bf 1.}  It is well known that Coulomb type potentials generate infrared divergences
(infrared catastrophe) . The problem has a long history starting in 1940 with  the fundamental works of
R. Feynman. Here, we construct a rigorous  solution of the
corresponding problem for a broad class of  cases.
Recall the operator function
\begin{equation} S(t,\tau,\varepsilon)=\exp(i t A_0)\exp(-i tA)\exp(i\tau A))\exp(-i\tau A_0), \label{3.1}
\end{equation}
which was mentioned in Introduction.
Recall also that the operators $A,\, A_0 , \A_1$ satisfy \eqref{1.1}. It follows from \eqref{3.1} that
\begin{equation}\frac{\partial}{\partial{t}}S(t,\tau,\varepsilon)=-i{\varepsilon}V(t)S(t,\tau,\varepsilon),\quad
\frac{\partial}{\partial{\tau}}S(t,\tau,\varepsilon)=i{\varepsilon}S(t,\tau,\varepsilon)V(\tau), \label{3.2}\end{equation}
where $t$ and $\tau$ belong to the {\it real axis} $\BR$ and
\begin{equation} V(t)=\exp(itA_0)A_1\exp(-itA_0),\quad S(t,t,\varepsilon)=I.\label{3.3}\end{equation}

{\bf 2.} Further in the sections, the operators $A,\, A_1,\,A_0$ are unknown. Hence,  we do not suppose that the 
operator $V(t)$ in \eqref{3.2} has the form  \eqref{3.3}. Therefore, in our  further considerations,
the operator $V(t)$ itself is called a perturbation operator.
Using the method of successive approximation and relation \eqref{3.2}, we obtain our next proposition.
\begin{Pn}\label{Proposition 3.1} Assume that  $V(t)$ is a self-adjoint, continuous and bounded operator function in the domain $1\leq |t| \leq T$.  Then, there exists such 
$\varepsilon _{T}>0$ that the series
\begin{equation} S(t,\tau,\varepsilon)=\sum_{p=0}^{\infty}S_{p}(t,\tau)\varepsilon^{p}\quad (S_0(t,\tau) \equiv I) \quad \label{3.4}\end{equation}
is convergent in the domains  $\tau=1, \,\, 1 \leq t\leq T$ and $t=-1, -T \leq \tau \leq-1$ for  $|\varepsilon| \leq \varepsilon _{T}$. \end{Pn}
Here, for $p\geq 0$ we have
\begin{align} & S_{p+1}(t,\tau)=-i\int_{\tau}^{t}V(t_1)S_p(t_1,\tau)dt_1=-i\int_{\tau}^{t}S_p(t,\tau_1)V(\tau_1)d\tau_1.\label{3.5}\end{align}

In the present article, we consider the operator function $V$  of the form
\begin{align} & V(t)=\frac{1}{t}B_{+}+u(t) \quad (t \geq 1),\quad V(t)=\frac{1}{t}B_{-}+u(t)\quad (t\leq -1), \label{3.6}
\end{align}
where $B_{\pm}$ are self-adjoint, bounded operators and $u(t)$ is a self-adjoint, continuous operator function such that
\begin{equation} \norm{u(t)}=O(|t|^{-\nu})\quad (\nu>1) \quad {\mathrm{for}} \quad  |t| \to  \infty.\label{3.7}\end{equation}
Relations \eqref{3.4}--\eqref{3.6} imply that
\begin{align}& S_1(t,1)=-i\left(B_{+}\ln(t)+\int_{1}^{t}u(s)ds\right)\quad {\mathrm{for}} \quad t \geq 1,  \label{3.8-}
\\ &
S_1(-1,\tau)=i\left(B_{-}\ln(|\tau|)+\int_{\tau}^{-1}u(s)ds\right)\quad {\mathrm{for}} \quad \tau \leq -1. \label{3.8}\end{align}

The following deviation factors were constructed for the perturbation operators $V(t)$ of the form \eqref{3.6}  in \cite{Sak}:
\begin{equation}W_{0}(t,\varepsilon)=t^{i\varepsilon{B_{+}}}\quad (t \geq 1),\quad W_{0}(t,\varepsilon)=(-t)^{i\varepsilon{B_{-}}} \quad (t \leq -1).
\label{3.9}\end{equation}
\begin{Rk} \label{RkNes}
Physically meaningful values of $t$ and $\tau$ are $t\geq 0$ and $\tau \leq 0$. Therefore, we will need 
the values of $V(t)$ and $W_0(t)$ for $|t|<1$ in further considerations.  We set
\begin{align}&
V(t)=u(t) \quad (|t|<1), \quad W_0(t)=I  \quad (|t| < 1).
\label{3.6+}\end{align}
Since we are interested in the asymtotics of $S(t,\tau,\ve)$ at infinity, relations \eqref{3.6+} are less important than
\eqref{3.6} and \eqref{3.9}.
\end{Rk}

Let us introduce the  operator function $S^{R}$ (regularized $S$):
\begin{align}& S^{R}(t,\tau,\varepsilon)=W_{0}(t,\varepsilon)S(t,\tau,\varepsilon)W_{0}^{-1}(\tau,\varepsilon) \label{3.10}\end{align}
It follows from \eqref{3.2},  \eqref{3.6} and \eqref{3.9}--\eqref{3.10} that
\begin{align}& \frac{\partial}{\partial{t}}S^{R}(t,\tau,\varepsilon)=-i{\varepsilon}U(t,\varepsilon)S^{R}(t,\tau,\varepsilon),\quad
S^{R}(t,t,\varepsilon)=I,\label{3.11}
\\ &
\frac{\partial}{\partial{\tau}}S^{R}(t,\tau,\varepsilon)=i{\varepsilon}S^{R}(t,\tau,\varepsilon)U(\tau,\varepsilon), \quad S^{R}(\tau,\tau,\varepsilon)=I,\label{3.12}\end{align}
where
\begin{equation}U(t,\varepsilon)=W_{0}(t,\varepsilon)u(t)W_{0}^{-1}(t,\varepsilon)\quad (t \in \BR).\label{3.13}\end{equation}
Clearly, 
we  have
\begin{align}&
S^R(0,0,\ve)=I \quad {\mathrm{and}} \quad S^{R}(t,\tau,\varepsilon)=S(t,\tau,\varepsilon), \quad U(t,\varepsilon)=u(t)
\label{3.13+}\end{align}
for $t,\tau \in [-1,1]$.
The next theorem easily follows from the relations of this section.
\begin{Tm}\label{Theorem 3.4}
Let the operator function $u(t)$ be bounded and continuous on  $\BR$. Then,
$S^{R}(t,\tau,\varepsilon)$ given by \eqref{3.11}--\eqref{3.13}
admits $($for $t\geq 1$ and $\tau \leq -1)$ the representation
\begin{equation} S^{R}(t,\tau,\varepsilon)=S^{R}(t,1,\varepsilon)S^R(1,-1,\varepsilon)S^{R}(-1,\tau,\varepsilon).\label{3.17}
\end{equation}
\end{Tm}
{\it Proof}.
Using multiplicative integrals (see appendix \ref{ApB}) and \eqref{3.11} we obtain
\begin{align} & S^{R}(t,\tau,\varepsilon)= \overset{t}{\overset{\curvearrowleft}{\underset{\tau}{\int}}}
e^{-i\varepsilon{U(s,\varepsilon)ds}}\qquad (t \geq \tau).\label{3.15}
\end{align}
Hence, for $t\geq 1$ and $\tau \leq -1$ we have
\begin{equation} S^{R}(t,\tau,\varepsilon)=\overset{t}{\overset{\curvearrowleft}{\underset{1}{\int}}}e^{-i\varepsilon{U(s,\varepsilon)ds}}
\overset{1}{\overset{\curvearrowleft}{\underset{-1}{\int}}}e^{-i\varepsilon{U(s,\varepsilon)ds}}\overset{-1}{\overset{\curvearrowleft}{\underset{\tau}{\int}}}
e^{-i\varepsilon{U(s,\varepsilon)ds}} .
\label{3.18-}
\end{equation}
Formula \eqref{3.17} immediately follows from \eqref{3.15} and \eqref{3.18-}.
$\Box$

Formula  \eqref{3.1} implies that 
the operators $S(t,\tau,\ve)$
are unitary for the case, where the self-adjoint operators $A_0$ and $A_1$ exist. In our general case, it follows from \eqref{3.6} and \eqref{3.9} that $W_0(t)$ is unitary. Hence, \eqref{3.13} implies that  $U(t,\ve)$
is self-adjoint. Now,  relation \eqref{3.15} yields the unitarity assertion:
\begin{Pn}\label{Proposition 3.5} The operators
\begin{equation}S^{R}(t,\tau,\varepsilon),\,\, S^{R}(t,1,\varepsilon),\,\, S^R(1,-1,\varepsilon),\,\, S^{R}(-1,\tau,\varepsilon),
\label{3.18}
\end{equation}
where $t\geq 1$ and $\tau\leq -1$, are unitary.\end{Pn}
In fact, it is easy to see that all the operators $S^R(t,\tau,\ve)$  are unitary (without the restriction $t\geq 1$ and $\tau\leq -1$). However,  only the unitarity
of the operators given in \eqref{3.18} is used further in the text.
\begin{Rk} Clearly, formula \eqref{3.17} $($under conditions of Theorem \ref{Theorem 3.4}$)$ may be rewritten as
\begin{equation} S^{R}(t,\tau,\varepsilon)=S^{R}(t,1,\varepsilon)S^R(1,0,\varepsilon)S^R(0,-1,\varepsilon)S^{R}(-1,\tau,\varepsilon),
\label{3.174}
\end{equation}
which is sometimes more convenient in the case of a special behaviour at zero. We also have
\begin{equation} S^{R}(t,0,\varepsilon)=S^{R}(t,1,\varepsilon)S^R(1,0,\varepsilon).
\label{3.172}
\end{equation}
All the factors in \eqref{3.174} and \eqref{3.172} are again unitary.
\end{Rk}

Next we  formulate the main result of this section, which follows from Theorem \ref{Theorem 3.4}, formula \eqref{3.13+}, Propositions \ref{Proposition 3.5} and \ref{Proposition 8.1}, and Corollary \ref{CyLeft}.
\begin{Tm}\label{Theorem 3.6} Let the conditions of Theorem \ref{Theorem 3.4} and relation \eqref{3.7} hold. Then, $S^{R}(t,\tau,\varepsilon)$ converges at each $\ve$
by norm $($to
the unitary operator  $\break S^{R}(+\infty,-\infty,\varepsilon))$ when $t$ tends to infinity and $\tau$ tends to minus infinity. Moreover, we have
\begin{equation} S^{R}(+\infty,-\infty,\varepsilon)=S^{R}(+\infty,1,\varepsilon)S(1,-1,\varepsilon)S^{R}(-1,-\infty,\varepsilon),
\label{3.22}
\end{equation}
where
\begin{equation}S^{R}(+\infty,1,\varepsilon)= \overset{+\infty}{\overset{\curvearrowleft}{\underset{1}{\int}}}
e^{-i\varepsilon{U(s,\varepsilon)ds}},\quad S^{R}(-1,-\infty,\varepsilon)=\overset{-1}{\overset{\curvearrowleft}{\underset{-\infty}{\int}}}
e^{-i\varepsilon{U(s,\varepsilon)ds}}.\label{3.23}\end{equation}\
\end{Tm}
\begin{Rk}\label{RkDef} The notion of the generalized scattering  operator was introduced in \cite{Sakh2} $($see also appendix \ref{ApA}$)$ and fruitfully used in \cite{Sakh3, Sakh1, Sakh10, Sakh10+, Sak}. In this section, we introduce a more general notion of the {\rm secondary generalized scattering  operator}. Namely, we do not require the
existence of the operators $A,\, A_0,\, A_!$. Under conditions of Theorem \ref{Theorem 3.6}, the operator $S^{R}(+\infty,-\infty,\varepsilon)$ is the secondary generalized scattering  operator.\end{Rk}
\begin{Rk}\label{Remark 3.8} Theorem \ref{Theorem 3.6} gives a positive answer  to Oppenheimer's question, which is cited in Remark \ref{Remark 1.1}.
\end{Rk}
Our next proposition is important for the theory but easily follows from Proposition \ref{Proposition 3.1} and formulas  \eqref{3.5} (for $p=0$), \eqref{3.8-} and \eqref{3.8}.
\begin{Pn}\label {Proposition 3.9} Under conditions of Proposition \ref{Proposition 3.1}  we have
 \begin{equation}V(t)=i\frac{d}{dt}S_1(t,1)\quad (t\geq 1), \qquad V(\tau)=-i\frac{d}{d\tau}S_1(-1,\tau) \quad (\tau\leq -1).\label{3.24}\end{equation}
Moreover, the perturbation operator $V(t)$ has the form  \eqref{3.6}
if and only if the first approximation $S_1$ has the form \eqref{3.7}.\end{Pn}
\begin{Rk}\label{Remark 3.10} Recall that in the classical quantum theory the perturbation  operator $V(t)$ is defined by  \eqref{3.3}. In our theory, the operators $A,\, A_1,\, A_0$ are unknown and may not exist. In our case, the operator $V(t)$ is recovered from the first approximation $S_1(t,\tau)$ using \eqref{3.24}.
\end{Rk}
{\bf 3.}
In a number of  theoretical and applied problems, the values of  $S_p$ $( p>1)$ are required in addition to the value $S_1$ of the first approximation  (see, e.g., \cite[f-la (47.52)]{AB}). 
Finding $S_p$ $(p>1)$ in a standard way is a complicated task. Below, we provide a simple solution for a wide range of problems of finding such $S_p$. Moreover, the expressions for such approximations  are written down  in an explicit form.   

First, we set
\begin{equation}\label{3.24+}
C_{1,0}=\int_{1}^{+\infty}u(s)ds,\quad Q_{1}(t)= -\int_{t}^{\infty}u(s)ds \quad (t\geq 1).
\end{equation}
The following theorem is valid.
\begin{Tm}\label{Theorem 3.14} Let the relations  \eqref{3.6},  \eqref{3.7} and \eqref{3.24+} be fulfilled. Then,
\begin{equation} S_{p+1}(t,1)=(-i)^{p+1}\left(\frac{B_{+}^{p+1}}{(p+1)!}\ln^{p+1}(t)+\sum_{k=0}^{p} C_{p+1,k}\ln^{k}(t)+Q_{p+1}(t)\right), \label{3.25} \end{equation}
where $p\geq 1$, $t \geq 1$ and
\begin{align} & \label{3.26}
C_{p+1,k}=B_{+} C_{p,k-1}/k \quad (1\leq k \leq p),
\\ &
C_{p+1,0}=\int_{1}^{+\infty}\left([ B_{+}/s+u(s)]Q_p(s)+u(s)\sum_{k=0}^{p-1}C_{p,k}\ln^k(s)\right)ds, \label{3.27}
\\ &
Q_{p+1}(t)=-\int_{t}^{+\infty}\left([ B_{+}/s+u(s)]Q_p(s)+u(s)\sum_{k=0}^{p-1}C_{p,k}\ln^k(s)\right)ds, \label{3.28}
\\ &
\norm{Q_{p}(t)}=O(t^{-\delta_p})\quad (\delta_p>0).\label{3.30}\end{align}\end{Tm}
It follows from \eqref{3.8-} and \eqref{3.24+} that \eqref{3.25}  holds for $p=0$. Next, the proof of  Theorem \ref{Theorem 3.14}  using  \eqref{3.5} and complete
mathematical induction is straightforward. 
According to the formulas \eqref{3.7} and \eqref{3.30}, the integrals in Theorem \ref{Theorem 3.14} are  absolutely converging.

Now, we set
\begin{equation}\label{3.30+}
C_{1,0}=\int_{-\infty}^{-1}u(s)ds,\quad Q_{1}(\tau)= -\int_{-\infty}^{\tau}u(s)ds \quad (\tau \leq -1).
\end{equation}

\begin{Tm}\label{Theorem 3.15} Let the conditions \eqref{3.6},  \eqref{3.7} and \eqref{3.30+} be fulfilled. Then,
\begin{equation} S_{p+1}(-1,\tau)=(i)^{p+1}\left(\frac{B_{-}^{p+1}}{(p+1)!}\ln^{p+1}(|\tau|)+\sum_{k=0}^{p} C_{p+1,k}\ln^{k}(|\tau|)+Q_{p+1}(\tau)\right), \label{3.31} \end{equation}
where $p\geq 1$, $\tau \leq -1$ and
\begin{align} & C_{p+1,k}= [C_{p,k-1}/k]B_{-} \quad (1 \leq k \leq p), \label{3.32}
\\ &
 C_{p+1,0}=\int_{-\infty}^{-1}\left(Q_p(s)[ B_{-}/s+u(s)]+[\sum_{k=0}^{p-1}C_{p,k}\ln^k(s)]u(s)\right)ds, \label{3.33}
 \\ &
  Q_{p+1}(\tau)=-\int_{-\infty}^{\tau}\left(Q_p(s)[ B_{-}/s+u(s)]+[\sum_{k=0}^{p-1}C_{p,k}\ln^k(s)]u(s)\right)ds,\label{3.34}
 \\ & 
 \norm{Q_{p}(\tau)}=O(|\tau]^{-\delta_p})\quad (\delta_p>0). \label{3.35}\end{align}\end{Tm}
It follows from \eqref{3.8} and \eqref{3.30+} that \eqref{3.31}  holds for $p=0$. Next, the proof of  Theorem \ref{Theorem 3.15}  using  \eqref{3.5} and complete
mathematical induction is straightforward. 
According to the formulas \eqref{3.7} and \eqref{3.35}, the integrals in Theorem \ref{Theorem 3.15} are  absolutely converging.

It is interesting that Heisenberg understood divergency problems encountered in the theory of elementary particles and apparently assumed that
 these problems could be overcome using his S-program  \cite{Hei}.
\section{Ultraviolet divergence}\label{sec4}
{\bf 1.} In physical studies of the ultraviolet case (see \cite{Sak} and references therein), a matrix function $d(L,q,\varepsilon)$ $(L>0, \, q\in \BR^4)$
is considered such that 
\begin{align}& d(L,q,\varepsilon)=1+\varepsilon\big(a_{1}(L,q)\big)+o(\varepsilon). \label{4.1}\end{align}
Here, $a_{1}(L,q)$ may be written down as an integral over the four dimensional sphere with radius $L$ (in spherical coordinates):
\begin{equation}
a_{1}(L,q)=-i\int_{0}^{L}\int_{0}^{2\pi}\int_{0}^{\pi}\int_{0}^{\pi}F(p,q)r^{3}\big(\sin^{2}\phi_{1}\big)\sin\phi_2d\phi_1d\phi_2d\phi_3dr,\label{4.2}
\end{equation}
where $F (p, q)$ is a rational matrix function, $p=[p_1,p_2,p_3,p_4],\,\, q=[q_1,q_2,q_3,q_4]$ $\break (p,q\in \BR^4)$ and
\begin{align}& p_1=r\cos\phi_1,\quad p_2=r\sin\phi_1\cos\phi_2, \label{4.3-}
\\ &
p_3=r\sin\phi_1\sin\phi_2\cos\phi_3, \quad
p_4=r\sin\phi_1\sin\phi_2\sin\phi_3 .\label{4.3}\end{align}
In the classical case, it is assumed that the limit of  $d(L,q,\varepsilon)$ $(L\to \infty)$ exists: $d(q,\varepsilon)=\lim_{L\to \infty}d(L,q,\varepsilon)$.
We will study the case, where the limit of $a_1(L,q)$ (for $L\to \infty$) does not exist. In that case we have ultraviolet divergence.  In particular,
formula \eqref{4.2} is actively used in section \ref{sec5}.

{\bf 2.} The results of section \ref{sec3} may be applied  in section \ref{sec4}, if we replace the variable $t$ by $L$.
Thus, the expression  $a_1(L,q)$ is a version of  the expression $S_1(t,\tau)$. The equality
\begin{equation}V(L,q)=i\frac{d}{dL}a_1(L,q). \label{4.3+} \end{equation}
is a version of relation \eqref{3.24}.  It follows from \eqref{4.2} and \eqref{4.3+} that
\begin{equation} V(L,q)=\int_{0}^{2\pi}\int_{0}^{\pi}\int_{0}^{\pi}F(P,q)L^{3}\big(\sin^{2}\phi_{1}\big)\sin\phi_2d\phi_1d\phi_2d\phi_3,\label{4.4}\end{equation}
where $P=[P_1,P_2,P_3,P_4]$ and
\begin{align}& P_1=L\cos\phi_1,\quad P_2=L\sin\phi_1\cos\phi_2, \nn
\\ &
P_3=L\sin\phi_1\sin\phi_2\cos\phi_3, \quad
P_4=L\sin\phi_1\sin\phi_2\sin\phi_3 .\nn\end{align}
It is known that the ultraviolet divergences may be removed using mass and charge renormalizations \cite{AB}. F.J. Dyson \cite{Dy}  stressed that it is important ``to prove the convergence in the frame of the theory". For a broad class of examples, it was done in our paper \cite{Sak} using deviation factors. The results of \cite{Sak} are valid only for the first approximation $a_1(q)$. In the present paper, we show that deviation factors and results of section 3 allow us to solve the divergence problems
(logarithmic divergence case) for all approximations, in other words for the scattering operator.

Interesting results on the asymptotic behaviour  of the Feynman integrals \eqref{4.2} are derived in \cite{Du} (see also some other papers and references
in \cite{AC}).
\begin{Ee}\label{Example 4.1} Let us consider the case where $a_1(L,q)$ is a scalar function and
 \begin{equation}a_{1}(L,q)=-i\left(\phi(q)\ln{L}+\int_{1}^{L}u(s,q)ds\right).
\label{4.5}\end{equation}\end{Ee}
Here, $\phi(q)$ and $u(L,q)$ are bounded, self-adjoint operators $($of multiplication by the functions of $q)$ and
\begin{equation} \norm{u(L, q)}=O(L^{-\nu}) \quad (\nu>1,\quad  L \geq 1).\label{4.6}\end{equation}
In the spirit of  R. Feynman's ``space-time approach" \cite{Fey1} and according to \eqref{4.3+} and \eqref{4.5}, we have
\begin{equation} V(L,q)=\frac{\phi(q)}{L}+u(L,q),\quad \big(B_{+}f\big)(q)=\phi(q)f(q).\label{4.7}\end{equation}
Using $V(L,q)$, we introduce the differential equation for the ultraviolet case in the form 
\begin{equation}\frac{d}{dL}S(L,q,\varepsilon)=-i{\varepsilon}V(L,q)S(L,q,\varepsilon) \quad  \big(S(1,q,\varepsilon)=1, \quad L\geq 1),\label{4.8}\end{equation}
where $V(L,q)$ is given by \eqref{4.7}.  The solution of \eqref{4.8} may be presented as a series:
\begin{equation}\nn S(L,q,\varepsilon)=1-i\varepsilon\int_{1}^{L}V(s,q)ds+...\end{equation}
The corresponding deviation factor $W_{0}(L,q,\varepsilon)$  has the form
\begin{equation}W_{0}(L,q,\varepsilon)=L^{i\varepsilon\phi(q)}.\label{4.9}\end{equation}
A regularized operator function $S^R$ is introduced  by the formula
\begin{equation}S^{R}(L,q,\varepsilon)=W_{0}(L,q,\varepsilon)S(L,q,\varepsilon) \label{4.10}\end{equation}
It follows from \eqref{4.8}--\eqref{4.10} that
\begin{equation}\frac{d}{dL}S^{R}(L,q,\varepsilon)=-i{\varepsilon}U(L,q,\varepsilon)S^{R}(L,q,\varepsilon),\quad
S^{R}(1,q,\varepsilon)=1,\label{4.11}\end{equation}
where
\begin{equation}U(L,q,\varepsilon)=W_{0}(L,q,\varepsilon)u(L,q)W_{0}^{-1}(L,q,\varepsilon),\quad L \geq 1.\label{4.12}\end{equation}
Using multiplicative integrals (see Appendix \ref{ApB}) and relation \eqref{4.11} we have
\begin{equation} S^{R}(L,q,\varepsilon)= \overset{L}{\overset{\curvearrowleft}{\underset{1}{\int}}}
e^{-i\varepsilon{U(s,q,\varepsilon)ds}},\quad L{\geq}1.\label{4.13}\end{equation}
Next, we  formulate the main result in this section.
\begin{Tm}\label{Theorem 4.2} Let conditions \eqref{4.5} and \eqref{4.6} be fulfilled. Then, \begin{equation}S^{R}(L,q,\varepsilon){\Longrightarrow}S^{R}(+\infty,q,\varepsilon)\quad 
{\mathrm{for}} \quad L{\to}+\infty,\label{4.14}
\end{equation}
where the symbol $\Longrightarrow$ denotes convergence by norm and
\begin{equation}S^{R}(+\infty,q,\varepsilon)= \overset{+\infty}{\overset{\curvearrowleft}{\underset{1}{\int}}}
e^{-i\varepsilon{U(s,q,\varepsilon)ds}}. \label{4.15}\end{equation}\end{Tm}
Here, we call $S^R(+\infty, q,\varepsilon)$ {\it the secondary generalized
scattering operator.}
\begin{Rk}\label{Rk4.3}  In a similar way, Theorem \ref{Theorem 4.2}  may be formulated and proved in the case of the matrix function $a_1$.
\end{Rk}
\section{The simplest concrete examples, \\ Feynman's integrals}\label{sec5}
Integrals of the form \eqref{4.2} are studied and their physical interpretation is given in the famous paper \cite{Fey1} by 
R.P. Feynman.
In the present section, we will illustrate our approach to the divergence problems (see sections \ref{sec3} and \ref{sec4})
by concrete physical examples of the form \eqref{4.2} from \cite{AB}.
Let us introduce the   Pauli  matrices $\sigma_{k}$ and $\alpha_{k}$:
\begin{equation}\alpha_k=\left(
                           \begin{array}{cc}
                             0 & \sigma_k \\
                             \sigma_k & 0 \\
                           \end{array}
                         \right), \quad k=1,2,3,\label{5.1}\end{equation}
where
\begin{equation}\sigma_1=\left(
                           \begin{array}{cc}
                             0 & 1 \\
                             1 & 0 \\
                           \end{array}
                         \right),\quad
\sigma_2=\left(
                           \begin{array}{cc}
                             0 & -i \\
                             i & 0 \\
                           \end{array}
                         \right),\quad
\sigma_3=\left(
                           \begin{array}{cc}
                             1 & 0 \\
                             0 & -1 \\
                           \end{array}
                         \right).\label{6.2}\end{equation}
The matrices  $\beta,\,\gamma_{\mu}$ and $\hat{p}$  are defined by the relations
\begin{equation} \beta=\left(
                         \begin{array}{cc}
                            I_2 & 0 \\
                           0 & -I_2 \\
                         \end{array}
                       \right),\quad
\gamma_{j}=\beta\alpha_{j},\quad (j=1,2,3),\quad \gamma_4=\beta,\quad
\hat{p}=\sum_{\mu=1}^4p_{\mu}\gamma_{\mu}.
\label{5.4}\end{equation}
where $I_k$ is the $k \times k$ identity matrix. Similar to $\hat p$, we have $\hat{q}=\sum_{\mu=1}^4 q_{\mu}\gamma_{\mu}$.
\begin{Ee}\label{Example 5.1} Let the first approximation $a_1(L,q,\mu)$ have the form \eqref{4.2}, where  $F(p,q,\mu)$ is a $4\times 4$ matrix function:
\begin{equation} F(p,q,\mu)=\frac{1}{(2\pi)^4p^2}\gamma_{\mu}\frac{i(\hat{q}-\hat{p})-mI_4}{(q-p)^2+m^2}
\gamma_{\mu},\label{5.5}\end{equation}
$p=[p_1,p_2,p_3,p_4],\, p^2=p_{1}^2+p_{2}^2+p_{3}^2+p_{4}^2,$  and $m$ is a constant $($usually it is a mass$)$.
Here, $J_{\mu}(L,q):=a_1(L,q,\mu)$ are the so called Feynman's integrals.\end{Ee}

We note that the integrals $J_{\mu}(L,q)$ play an important role in electron collision problems \cite{AB}. 
In \cite{Sak}, we derived the equalities
 \begin{align}\nn
 J_{\mu}(L,q)=&\frac{1}{(2\pi)^4}\left(i  m \pi^{2}(2\ln{L}-1)I_{4}-\frac{\pi^{2}}{2}\gamma_{\mu}\hat{q}\gamma_{\mu}
 -im\pi^{2}(\ln(B(q))I_{4}\right)
 \\ & \label{5.6}
 +R_{\mu}(L,q) \quad {\mathrm{for}} \,\, \mu=1,2,3;
 \\ \nn
J_{4}(L,q)=&-\frac{1}{(2\pi)^4}\left(i m\pi^{2}(2\ln{L}-1)I_{4}+\frac{\pi^{2}}{2}\gamma_{4}\hat{q}\gamma_{4}
 -i m\pi^{2}(\ln(B(q))I_{4}\right)
\\ & 
 +R_{4}(L,q),
\label{5.7}\end{align}
where $B$ and $R_{\mu}$ satisfy relations
\begin{align}& \label{5.8}  \ln(B(q))=\int_{0}^{1}\ln\big(\ell(p,v)-q^{2}v^{2}\big)dv, \quad \ell(p,v)=(p^2+m^2)v;
\\ &
\lim_{L{\to}+\infty}R_{\mu}(L,q)=0\quad (\mu=1,2,3,4).
\label{5.8+}\end{align}
It follows from \eqref{3.24}  that
\begin{equation}V(L,q,\mu)=i\frac{d}{dL}a_1(L,q,\mu).\label{5.10}\end{equation}
Hence, taking into account \eqref{5.6}--\eqref{5.8}  we obtain:
\begin{equation} V(L,q,\mu)=\frac{B_{+}}{L}+u(L,q,\mu),\label{5.11}\end{equation}
where
\begin{equation} B_{+}=\frac{m}{2(2\pi)^{2}},\quad u(L,q,\mu)=\frac{d}{dL}R_{\mu}(L,q).\label{5.12}\end{equation}
In the case of Example 5.1,  $\frac{d}{dL}R_{\mu}(L,q)$ is rational with respect to $L$ and admits Laurent series
representation at $L=\infty$.  Hence, taking into account \eqref{5.8+} one can see that $R_{\mu}(L,q)$ admits
Laurent series representation $\sum_{k\geq 1} C_k(q)/L^k$. In this way, one obtains 
$$\frac{d}{dL}R_{\mu}(L,q)=\sum_{k\geq 1} \big(-kC_{k}(q)\big)\big/L^{k+1}.$$
Now, in view of \eqref{5.12}, it is easy to show that
 condition  \eqref{4.6} is fulfilled with $\nu=2$.
 
 Finally, let us rewrite formula \eqref{4.2} in the form
\begin{align}\nn a_{1}(L,q,\mu)=&-i\int_{1}^{L}\int_{0}^{2\pi}\int_{0}^{\pi}\int_{0}^{\pi}F(p,q,\mu)r^{3}\big(\sin^{2}\phi_{1}\big)\sin\phi_{2}d\phi_1d\phi_2d\phi_3dr 
\\ \nn &
-i\psi_{\mu}(q),\end{align}
where 
$$\psi_{\mu}(q)=\int_{0}^{1}\int_{0}^{2\pi}\int_{0}^{\pi}\int_{0}^{\pi}F(p,q,\mu)r^{3}\big(\sin^{2}\phi_{1}\big)\sin\phi_{2}d\phi_1d\phi_2d\phi_3dr. $$
The deviation factor $W_{0}(L,q,\ve,\mu)$ is defined by the formula 
\begin{equation}\label{5.12+} W_{0}(L,q,\ve,\mu)=L^{i\varepsilon\phi}e^{i \varepsilon\psi_{\mu}(q)}, \quad \phi=\frac{m}{8\pi^2}.\end{equation}
Using formulas \eqref{4.10}--\eqref{4.13}, we can show that  Theorem \ref{Theorem 4.2} and Remark \ref{Rk4.3} are valid in the case of Example \ref{Example 5.1}.

\begin{Ee}\label{Example 5.2} Let the first approximation $a_1(L,q)$ have the form \eqref{4.2} with
\begin{equation} F(p,q)=\frac{p_{\sigma}p_{\tau}}{(p^2-2pq+\ell(q))^3} \qquad(\ell(q)>q^2),\label{5.13}\end{equation}
where $\sigma$ and $\tau$ are fixed positive integers $(1\leq \sigma, \tau \leq 4)$ and $pq=\sum_{k=1}^4p_kq_k$.
\end{Ee}

The following expression for $a_1(L,q)$ was obtained in \cite[(47.10)]{AB}:
\begin{equation}\label{5.14}
a_1(L,q)
=\frac{i\pi^2}{4}\delta_{\sigma\tau}\left(\ln \frac{L^2}{\ell(q)-q^2}-\frac{3}{2}\right)+\frac{i\pi^2}{2}\frac{q_{\sigma}q_{\tau}}
{\ell(q)-q^2}+iR(L,q), \end{equation}
where $\delta_{\sigma\tau}$ is the Kronecker-delta and
\begin{equation}\lim_{L{\to}+\infty}R(L,q)=0\qquad (R(L,q)\in \BR) .\label{5.15}\end{equation}
Using \eqref{5.10} and \eqref{5.15} we get
\begin{equation} V(L,q)=\frac{B_{+}}{L}+u(L,q),\label{5.18}\end{equation}
where 
\begin{equation} B_{+}=\phi,\quad u(L,q)=\frac{d}{dL}R(L,q),\quad \phi=\frac{\pi^{2}}{2}.\label{5.19}\end{equation}
Here, in a way similar to Example \ref {Example 5.1}  one may show that condition  \eqref{4.6} is fulfilled for $\nu=2$.
Similar to Example \ref {Example 5.1}, we   rewrite \eqref{4.2} in the form
\begin{equation}\label{5.29} a_{1}(L,q)=-i\int_{1}^{L}\int_{0}^{2\pi}\int_{0}^{\pi}\int_{0}^{\pi}F(p,q)r^{3}\big(\sin^{2}\phi_{1}\big)
\sin\phi_{2}d\phi_1d\phi_2d\phi_3dr -i\psi(q),\end{equation}
where 
\begin{align}\label{5.30}&
\psi(q)=\int_{0}^{1}\int_{0}^{2\pi}\int_{0}^{\pi}\int_{0}^{\pi}F(p,q)r^{3}\big(\sin^{2}\phi_{1}\big)\sin\phi_{2}d\phi_1d\phi_2d\phi_3dr .
\end{align}
In view of  \eqref{4.1}, \eqref{5.19}  and \eqref{5.29}, the deviation factor $W_{0}(L,q, \varepsilon)$ is defined by the formula  
\begin{equation}W_{0}(L,q,\ve)=L^{i\varepsilon\phi}e^{i\varepsilon\psi(q)},\end{equation}
where $\psi$ and $\phi$  are given by \eqref{5.13}, \eqref{5.30} and \eqref{5.19}, respectively.
Using again formulas \eqref{4.10}--\eqref{4.13}, we can show that  Theorem \ref{Theorem 4.2}  is valid in the case of Example \ref{Example 5.2}.


\begin{Ee}\label{Example 5.3} Let the first approximation $a_1(L,q)$ have the form \eqref{4.2} where
\begin{equation} F(p,q)=\frac{p_{\sigma}}{(p^2-2pq+\ell(q))^2}  \qquad(\ell(q)>q^2).\label{5.20}\end{equation}
\end{Ee}
The following expression for $a_1(L,q)$ was obtained in \cite[(47.12)]{AB}:
\begin{equation}a_1(L,q)=
i\pi^{2}q_{\sigma}\left(\ln \frac{L^2}{\ell(q)-q^2}-\frac{3}{2}\right)+iR(L,q).\label{5.21} \end{equation}
where
\begin{equation}\lim_{L{\to}+\infty}R(L,q)=0 \qquad (R(L,q)\in \BR).\label{5.22}\end{equation}
Hence,  we obtain
\begin{equation} V(L,q)=\frac{B_{+}}{L}+u(L,q),\label{5.24}\end{equation}
where
\begin{equation} B_{+}=2\pi^{2}q_{\sigma}=\phi .\label{5.25}\end{equation}
In view of  \eqref{4.1}, \eqref{5.29}  and \eqref{5.25}, the deviation factor $W_{0}(L,q, \varepsilon)$ is defined by the formula  
\begin{equation}W_{0}(L,q,\ve)=L^{i\varepsilon\phi}e^{i\varepsilon\psi(q)},\end{equation} 
where $\psi$ and $\phi$  are given by \eqref{5.30}, \eqref{5.20} and \eqref{5.25}, respectively.

In the case of Example \ref{Example 5.3}  condition  \eqref{4.5} is fulfilled and
Theorem \ref{Theorem 4.2}  is again valid.
Some other concrete examples are contained in the article \cite{Sak}.

\section{Conclusion}\label{sec6}

Feynman's theory assumes that some information about the scattering operator (namely, its first approximation) is given, and does not assume the existence of the 
perturbed operator $A$, unperturbed operator $A_0$  and perturbation operator $A_1$. According to \cite{Fey0},
``in a scattering problem this over-all view of the  complete scattering process is similar to the S-matrix view-point
of Heisenberg".

In the present article, we use the well-known equation \eqref{1.3}. In the classical case (see, e.g., \cite{AB}),
the operator function  $V(t)$  in \eqref{1.3} is defined via \eqref{1.4} using the operators $A,\,A_0$ and  $A_1$.
Here, we define the operator function $V(t)$ in \eqref{3.24} using the first approximation of the scattering operator.
Similar to the Feynman's theory, we do not use the operators  $A$ and $A_1$ and $A_0$. Those operators are not given, and, may be, do not exist.
Let us discuss some of the useful developments in this paper.

{\bf 1.}  The first approximation of the scattering operator
may tend to infinity when one of the parameters tends to infinity, and
Feynman's theory gives a numerically precise method for discarding terms tending to infinity.
A rigorous justification of this method has long been an open problem.

\emph{ In the present  article, we give  a  rigorous approach to the divergence problems in QED for logarithmic divergences.}

{\bf 2.}  In  our paper  \cite{Sak}, a rigorous procedure for regularizing  the first  approximation of the scattering operator is given for 
varoius concrete examples and so called deviation factors are used for this purpose (see Appendix \ref{ApA}).
In the present   article, the deviation factors 
 are used for the  general case of  logarithmic divergences.
Moreover,  not only the first but all  approximations are derived  in the explicit form (see\eqref{3.25}).

{\it Thus, the presented article proposes a new and fruitful approach to the realization  of the Heisenberg's S-program.
In the framework of this approach, we solve the problem of the rigorous treatment of the divergencies.
Moreover, the divergences generate the physically meaningful deviation factors, and so the divergencies
are  physically meaningful  as well.}
In addition to the examples considered here, the self-energy examples with logarithmic divergence
from the seminal paper \cite{Fey1} (see also \cite{Cha}) will be considered in our next paper.
We also assume that the ideas, which are presented here, are suitable for other types of divergences  
(e.g., for the cases of divergencies appearing in  Example 4.9 and section 5 of our paper \cite{Sak}).

\vspace{0.5em}

\noindent {\bf Acknowledgement}

The author is grateful to A.L. Sakhnovich for fruitful discussions and his help in the
preparation of the manuscript.

\appendix

\section{Generalized wave operators  and \\ deviation factors}\label{ApA}
Wave operators play an essential role in many problems of mathematical physics.
However, the wave operators do not exist, when the initial and/or final states of the system may not be regarded as free.
In these cases, one has to consider the {\it generalized wave operators}  (see, e.g.,  \cite{Do, Sakh2,  Sakh3, Sakh10, Sakh10+, Sak}).

Let $A$ and $A_0$ be linear self-adjoint (not necessary bounded) operators acting in some Hilbert space $H$. 
Denote the absolutely continuous subspace  of the operator $A$ (i.e.,  the subspace corresponding
to the absolutely continuous spectrum of $A$) by $G$ and the absolutely continuous subspace  of  $A_0$ by $G_0$.
Below,  the notions  of the generalized wave operators $W_{\pm}(A,A_{0})$  mapping $G_0$ into $G$ and of the unitary deviation factor (operator function) $W_{0}(t)$
acting in $G_0$ are introduced \cite{Sakh2,Sakh3, Sak}.

 \begin{Dn}\label{Definition 7.1}
 Operator functions $W_{\pm}(A,A_{0})$ and  $W_{0}(t)$ are called the generalized wave operators and a deviation factor, respectively, if
the following conditions are fulfilled:

1. The limits
\begin{equation}W_{\pm}(A,A_{0})
=\underset{t\to\pm\infty}{s-\lim}\, [e^{iAt}e^{-iA_{0}t}W_{0}^{-1}(t)]
\label{7.1}\end{equation}
exist in the sense of strong convergence.

2. The operators $W_{0}(t)$ and  $W_{0}^{-1}(t)$  acting in  $G_0$
are unitary for all $t$ and
\begin{equation}\lim_{t{\to}\pm\infty}W_{0}(t+\tau)W_{0}^{-1}(t)=I_{G_0}\quad {\mathrm{for\, all}}\quad \tau=\overline{\tau},
\label{7.2}\end{equation}
where $I_{G_0}$ is the identity operator  in $G_0$.

3. The following commutation relations  hold for the arbitrary values of $t$ and $\tau$:
\begin{equation}W_{0}(t)A_{0}=A_{0}W_{0}(t),\quad
W_{0}(t)W_{0}(t+\tau)=W_{0}(t+\tau)W_{0}(t).
\label{7.3}\end{equation}
\end{Dn}
If  $W_{0}(t)\equiv I_{G_0}$, then the operators $W_{\pm}(A,A_{0})$ are usual wave
operators.
\begin{Dn}\label{Definition 7.2}The generalized scattering operator $S(A,A_0)$
is defined by the formula:
\begin{equation}S(A,A_0)=W^{*}_{+}(A,A_0)W_{-}(A,A_0),\label{7.4}
\end{equation}
where $W_{\pm}(A,A_{0})$ are generalized wave operators.\end{Dn}
Here, the operator $W^{*}_{+}$ is adjoint to $W_{+}$.
The operator $S(A,A_0)$ is a unitary mapping of $G_0$  onto itself and
\begin{equation}A_{0}S(A,A_0)=S(A,A_0)A_{0}.\label{7.5}\end{equation}

In our theory, the operators $A$, $A_1$ and $A_0$ do not exist. Hence, the generalised wave operators do not exist too.
In this case, we define the deviation factor with the help of the first approximation of  $S$ (see section \ref{sec3}).
\begin{Rk}\label{Remarl 7.4} If we generalize $W_0(t)$ for the case where $A$, $A_1$ and $A_0$ do not exist,
the following properties of  the deviation factor from Definition~\ref{Definition 7.1} seem  the most important$\,:$ \\
1. The operators $W_{0}(t)$ and  $W_{0}^{-1}(t)$ are
unitary for all $t$ and the limits
\begin{equation}\underset{t{\to}\pm\infty}{s-\lim}\, W_{0}(t+\tau)W_{0}^{-1}(t)=I\quad (\tau=\overline{\tau}).
\label{7.6}\end{equation}
exist in the sense of strong convergence.\\
2. The commutation relation
\begin{equation}
W_{0}(t)W_{0}(t+\tau)=W_{0}(t+\tau)W_{0}(t)
\label{7.7}\end{equation}
holds for arbitrary values of $t$ and $\tau$.

In particular, the condition 1.  is fulfilled for the generalized  deviation factor in the case of the logarithmic divergence  $($see the present work and our paper \cite{Sak}$)$.\end{Rk}

Generalized wave operators were also used  in \cite{Kul}.  Interesting papers on the infrared case \cite{Gom, Kap}
are closely related to the Faddeev-Kulish construction \cite{Kul}.

\section{Multiplicative integrals}\label{ApB}
In our article, multiplicative integrals (see, e.g.,  \cite{DF,Gan, Pot}) are actively used.
In this appendix, we present the corresponding basic definitions and some assertions, which we need in the main text.
The right and left multiplicative integrals $\overset{b}{\overset{\curvearrowright}{\underset{a}{\int}}}e^{F(t)dt}$ and $\overset{b}{\overset{\curvearrowleft}{\underset{a}{\int}}}e^{F(t)dt}$
are defined, respectively, by the formulas
\begin{align}&
\overset{b}{\overset{\curvearrowright}{\underset{a}{\int}}}e^{F(t)dt}=\lim_{\max{\Delta_j}{\to}0}\big[e^{F(t_1)\Delta_1}e^{F(t_2)\Delta_2} \ldots e^{F(t_n)\Delta_n}\big],\label{8.1}
\\ &
\overset{b}{\overset{\curvearrowleft}{\underset{a}{\int}}}e^{F(t)dt}=\lim_{\max{\Delta_j}{\to}0}\big[e^{F(t_n)\Delta_n}e^{F(t_{n-1})\Delta_{n-1}}\ldots e^{F(t_1)\Delta_1}\big].\label{8.2}
\end{align}
Here $-\infty<a=t_0<...<t_n=b<\infty$, $\,\,\Delta_j=t_j-t_{j-1}$ and $\,F(t)$ is a continuous operator function. When $a \to -\infty$ or $b \to +\infty$ and the 
corresponding limits exist we obtain improper multiplicative integrals. For the right multiplicative integrals we have
\begin{align} &\overset{b}{\overset{\curvearrowright}{\underset{a}{\int}}}e^{F(t)dt}{\Longrightarrow} \overset{b}{\overset{\curvearrowright}{\underset{-\infty}{\int}}}e^{F(t)dt}
\,\,\,\, (a{\to} -\infty),\quad \overset{b}{\overset{\curvearrowright}{\underset{a}{\int}}}e^{F(t)dt}{\Longrightarrow} \overset{\infty}{\overset{\curvearrowright}{\underset{a}{\int}}}e^{F(t)dt}
\,\, \,\,(b\to \infty).
\label{8.3-}\end{align}
where the symbol $\Longrightarrow$ denotes convergence by norm. Similar notations are used for the left improper multiplicative integrals.
 \begin{Pn}\label{Proposition 8.1}If the operator function $F(t)$ is continuous and
 \begin{equation}\int_{a}^{\infty}\norm{F(t)}dt<\infty,\label{8.3}\end{equation}
 then the corresponding left multiplicative integrals converge by norm for $b$ tending to infinity
\begin{equation} \overset{b}{\overset{\curvearrowleft}{\underset{a}{\int}}}e^{F(t)dt}{\Longrightarrow} \overset{\infty}{\overset{\curvearrowleft}{\underset{a}{\int}}}e^{F(t)dt}
\qquad (b\to  \infty).
\label{8.4}\end{equation}
\end{Pn}
Relation \eqref{8.4} easily  follows from the inequality
\begin{equation} \Big\| \overset{t}{\overset{\curvearrowleft}{\underset{\tau}{\int}}}e^{F(t)dt}\Big\| \leq \exp\left(\int_{\tau}^{t}\norm{F(t)}dt\right), \quad
{\mathrm{where}} \quad  \tau<t.\label{8.5}\end{equation}
However, in spite of being easy to derive, Proposition \ref{Proposition 8.1} proves quite useful in this work.
\begin{Cy}\label{CyLeft} 
If the operator function $F(t)$ is continuous and
 \begin{equation}\int_{-\infty}^{b}\norm{F(t)}dt<\infty,\nonumber \end{equation}
 then the corresponding left and right multiplicative integrals converge by norm for $a\to -\infty :$
\begin{equation} 
\overset{b}{\overset{\curvearrowleft}{\underset{a}{\int}}}e^{F(t)dt}{\Longrightarrow} \overset{b}{\overset{\curvearrowleft}{\underset{-\infty}{\int}}}e^{F(t)dt},
\qquad
\overset{b}{\overset{\curvearrowright}{\underset{a}{\int}}}e^{F(t)dt}{\Longrightarrow} \overset{b}{\overset{\curvearrowright}{\underset{-\infty}{\int}}}e^{F(t)dt}
\qquad (a \to  -\infty).
\label{8.6}\end{equation}
\end{Cy}
\section {Time infinity}\label{ApC}
{\bf 1.}  Recall that the generalized wave operators have been discussed in Appendix \ref{ApA}.
In the simplest classical case, we assume that the self-adjoint operators $A$ and $A_0$ exist, $G=G_0=H$,
$W_0(t)=I_H$ and the  wave operators  $W_{\pm}(A,A_0)$  and $W_{\pm}(A_0,A)$ given by
\eqref{7.1} (where $W_0(t)=I_H$) exist as well.  It follows from \eqref{9.1} that 
\begin{equation} W_{\pm}(A_0,A)=W_{\pm}^{-1}(A,A_0)=W_{\pm}^{*}(A,A_0).\label{9.3} \end{equation}
For some initial vector $\psi_{0}\in H$, we set 
\begin{equation} \psi_{+}=W_{+}(A_0,A)\psi_0,\quad\psi_{-}=W_{-}(A_0,A)\psi_0.\label{9.1}\end{equation} 
According to \eqref{9.3} and  \eqref{9.1}, we have
\begin{equation} \psi_{+}=W_{+}(A_0,A)W_{-}(A,A_0)\psi_{-}=W_{+}^{*}(A,A_0)W_{-}(A,A_0)\psi_{-}\label{9.2} \end{equation}
or
\begin{equation} \psi_{+}=S(A,A_0)\psi_{-},\label{9.5} \end{equation}
where $S(A,A_0)$ is the scattering operator.

{\bf 2.}  Next we assume 
that  $G=G_0=H$ and the generalized wave operators $W_{\pm}(A,A_0)$  and $W_{\pm}(A_0,A)$ given by \eqref{7.1} (where $W_0(t)$
is not necessarily the identity operator) exist. Similar  to \cite[Ch. XI, Sect. 3]{ReeS} (and under the
corresponding conditions), one may show that
\begin{equation} W_{\pm}(A_0,A)=W_{\pm}^{-1}(A,A_0)=W_{\pm}^{*}(A,A_0).\label{9.8} \end{equation}
We set
\begin{equation} \phi_{+}=W_{+}(A_0,A)\psi_0,\quad\phi_{-}=W_{-}(A_0,A)\psi_0.\label{9.6}\end{equation} 
Similar to \eqref{9.2}, it follows  from \eqref{9.8} and \eqref{9.6} that
\begin{equation} \phi_{+}=W_{+}(A_0,A)W_{-}(A,A_0)\phi_{-}=W_{+}^{*}(A,A_0)W_{-}(A,A_0)\phi_{-}.\label{9.7} \end{equation}
Thus, taking into account \eqref{7.4} we obtain
\begin{equation} \phi_{+}=S(A,A_0)\phi_{-},\label{9.10} \end{equation}
where $S(A,A_0)$ is the generalized scattering operator.
For $t\to +\infty$ and $ \tau \to-\infty$ (according to \eqref{7.1} and \eqref{9.10}), we have
\begin{equation}\phi_{+}{\sim}
\big(e^{iAt}e^{-iA_{0}t}W_{0}^{-1}(t)\big)^{*}\big(e^{iA\tau}e^{-iA_{0}\tau}W_{0}^{-1}(\tau)\big)\phi_{-},
\label{9.11}\end{equation}
where the relation $ F(t,\tau)\sim G(t,\tau)$ means that
\begin{equation} s-\lim \big(F(t,\tau)-G(t,\tau)\big)=0\qquad ( t{\to}+\infty,\, \tau{\to}-\infty).\label{9.11+}\end{equation} 

Setting 
\begin{equation}\psi_{+}(t)=W_{0}^{-1}(t)\phi_{+},\quad \psi_{-}(\tau)=W_{0}^{-1}(\tau)\phi_{-},\label{9.13}\end{equation}
and recalling that the operators $W_0(t)$  are unitary, we derive
from \eqref{9.11}  that
\begin{equation}\psi_{+}(t){\sim}\big(e^{iAt}e^{-iA_{0}t}\big)^{*}\big(e^{iA\tau}e^{-iA_{0}\tau}\big)\psi_{-}(\tau)\quad {\mathrm{for}} \quad t{\to}+\infty,\, \tau{\to}-\infty.
\label{9.12}\end{equation}
Clearly,  relations \eqref{9.13} and \eqref{9.12}  are  also valid in the classical case, where $W_0(t)=I_H$.

{\bf 3.} Finally, let us consider the secondary generalized case. For this purpose, we use the relations \eqref{3.4}--\eqref{3.10}.
For some initial vector $\psi_{0}\in H$, we set
\begin{equation} \phi_{+}=s-\lim \, S^{R}(t,0,\ve)\psi_0,\quad \phi_{-}=s-\lim\, \big(S^{R}(0,\tau,\ve)\big)^*\psi_0
 \label{9.14}\end{equation}
for $t{\to}+\infty$ and $\tau{\to}-\infty$. It follows from the considerations of section \ref{sec3} that $S^R$ is a unitary operator.
Hence,  using \eqref{9.14} we derive
\begin{equation} \phi_{+}=s-\lim\big(S^{R}(t,0,\ve)S^{R}(0,\tau,\ve)\big)\phi_{-}\quad (t{\to}+\infty,\quad \tau{\to}-\infty). \label{9.15} \end{equation}
Taking into account \eqref{3.11}--\eqref{3.13}, we obtain
\begin{equation} \big(S^{R}(t,0,\ve)\big)^*=S^{R}(0,t,\ve),\quad S^{R}(t,0,\ve)S^{R}(0,\tau,\ve)=S^{R}(t,\tau,\ve).\label{9.16}\end{equation}
It follows from \eqref{9.15} and \eqref{9.16} that
\begin{equation}\psi_{+}(t) \sim 
S(t,\tau)\psi_{-}(\tau) \quad {\mathrm{for}} \quad t{\to}+\infty,\, \, \tau{\to}-\infty,
\label{9.17} \end{equation}
where
\begin{equation}\psi_{+}(t)=W_{0}^{-1}(t)\phi_{+},\quad \psi_{-}(\tau)=W_{0}^{-1}(\tau)\phi_{-}. \label{9.18}\end{equation}
\begin{Rk}
It would be very interesting to check relations \eqref{9.17}, \eqref{9.18} experimentally.
\end{Rk}

\end{document}